\documentclass[twocolumn]{emulateapj}

\slugcomment{}

\shorttitle{The Intrinsic Scatter along the main sequence of star-forming galaxies}
\shortauthors{Guo et al.}

\begin{document}

\title{The Intrinsic Scatter along the Main Sequence of Star-Forming Galaxies at ${z \sim 0.7}$}

\author{Kexin~Guo and Xian~Zhong\ Zheng}
\affil{Purple Mountain Observatory, Chinese Academy of Sciences, 2 West-Beijing Road, Nanjing 210008, P.R.China; kxguo,~xzzheng@pmo.ac.cn}

\author{Hai~Fu}
\affil{Department of Physics and Astronomy, University of Iowa, Van Allen Hall, Iowa City, IA 52242, USA; hai-fu@uiowa.edu}

\begin{abstract}
A sample of 12\,614 star-forming galaxies (SFGs) with stellar mass $>$10$^{9.5}M_\odot$ between $0.6<z<0.8$ from COSMOS is selected to study the intrinsic scatter of the correlation between star formation rate (SFR) and stellar mass. We derive SFR from ultraviolet (UV) and infrared (IR) luminosities. A stacking technique is adopted to measure IR emission for galaxies undetected at 24\,$\micron$. 
We confirm that the slope of the mass-SFR relation is close to unity. 
We examine the distributions of specific SFRs (SSFRs) in four equally spaced mass bins from 10$^{9.5}\,M_\odot$ to 10$^{11.5}\,M_\odot$. Different models are used to constrain the scatter of SSFR for lower mass galaxies that are mostly undetected at $24\,\micron$.
The SFR scatter is dominated by the scatter of UV luminosity and gradually that of IR luminosity at increasing stellar mass. We derive SSFR dispersions of 0.18, 0.21, 0.26 and 0.31\,dex with a typical measurement uncertainty of $\lesssim 0.01$\,dex for the four mass bins. Interestingly, the scatter of the mass-SFR relation seems not constant in the sense that the scatter in SSFR is smaller for SFGs of stellar mass $<10^{10.5}\,M_\odot$. If confirmed, this suggests that the physical processes governing star formation become systematically less violent for less massive galaxies. 
The SSFR distribution for SFGs with intermediate mass 10$^{10}-10^{10.5}$\,$M_\odot$ is characterized by a prominent excess of intense starbursts in comparison with other mass bins. We argue that this feature reflects that both violent (e.g., major/minor mergers) and quiescent processes are important in regulating star formation in this intermediate mass regime. 
\end{abstract}


\section{Introduction} \label{sec:intro}

Deep multi-wavelength extragalactic surveys revealed a tight correlation between star formation rate (SFR) and stellar mass for star-forming galaxies \citep[SFGs;][]{Noeske07a,Elbaz07,Daddi07a}. While the correlation, namely the {main sequence}, has been convincingly established from the local universe \citep{Brinchmann04} to the intermediate-redshift \citep{Pannella09,Peng10,Rodighiero10,Oliver10,Karim11} and high-redshift universe \citep{Stark09,Lee12,Papovich11,Reddy12}, more attention has now turned to exploring the normalization, slope and scatter of the relation $\log {\rm SFR} \propto \alpha \log M+\beta$ over cosmic time \citep[e.g.,][]{Whitaker12}.
The mass-SFR relation reaches its maximal normalization at $z$=2$-$3 \citep{Karim11} where the cosmic SFR density peaks \citep{Wilkins08}, and remains roughly constant \citep[e.g.,][]{Gonzalez10} or rises weakly out to $z\sim 7$ once contamination from nebular line emission is accounted for \citep[e.g.,][]{SD10, Gonzalez12, Stark13}. The slope of the main sequence is reported to be $\alpha$=$\sim$ 0.6$-$1 and is perhaps a function of redshift \citep[e.g.,][]{Pannella09,Karim11,Whitaker12}. Determination of the slope is affected by sample selection \citep{Stringer11,Salmi12} and observational biases \citep[e.g., extinction correction to SFR indicator;][]{Wuyts11a}. Also, the measured scatter in the main sequence is contributed by both observational uncertainties and the intrinsic scatter in specific SFR (SSFR=SFR/Mass). It is debatable whether the scatter depends on the stellar mass of SFGs, although a constant value of $\la0.3$\,dex is often suggested \citep[][but also see \citealt{Whitaker12}]{Noeske07a,Rodighiero11,Sargent12}.

The characteristics of the mass-SFR relation are linked to the physical processes regulating galaxy formation and evolution \citep{Dutton10,Hopkins10b,Leitner12}. Star formation in galaxies is believed to be regulated in general by both gas accretion from the cosmic web and feedback against gas cooling \citep[e.g.,][]{Bouche10}. The rapid decline of the normalization from $z \sim 2$ to $z=0$ is largely due to gas exhaustion \citep[e.g.,][]{Noeske07b,Zheng07a,Daddi07a, Magdis12, Tacconi13}, whereas the slope may be shaped by feedback and the intrinsic scatter is mostly induced by fluctuations in gas accretion rate and star formation efficiency. The intrinsic scatter in SSFR is correlated with gas mass fractions in galaxies \citep{Magdis12,Saintonge12}, galaxy environment \citep{Blanton09,Patel11}, interacting/merging processes \citep[][]{Lotz08,Jogee09}, morphologies \citep{Wuyts11b,Bell12}, and maybe galaxy mass \citep[e.g.,][]{Lee12}. 

Here we investigate the dependence of the scatter in the mass-SFR relation on galaxy stellar mass by reducing observational uncertainties and biases as much as possible. Such a study requires a large sample of well-defined SFGs and unbiased SFR estimation for both low-mass and high-mass galaxies.  
The Cosmic Evolution Survey \citep[COSMOS;][]{Scoville07} provides an ideal multi-wavelength data set. 
%
In \S \ref{sec:data}, we describe the data and galaxy sample. We present details of SFR estimation in \S \ref{sec:sfr} and our analysis of the mass-SFR relation in \S \ref{sec:ssfr}. Finally, we discuss and summarize our results in \S \ref{sec:discussion}. All magnitudes are given in the AB system. Throughout this work, we adopt a cosmology with $H_0=70$\,km\,s$^{-1}$\,Mpc$^{-1}$, $\Omega_{\rm M}=0.3$, and $\Omega_{\Lambda}=0.7$.

\begin{figure}
 \plotone{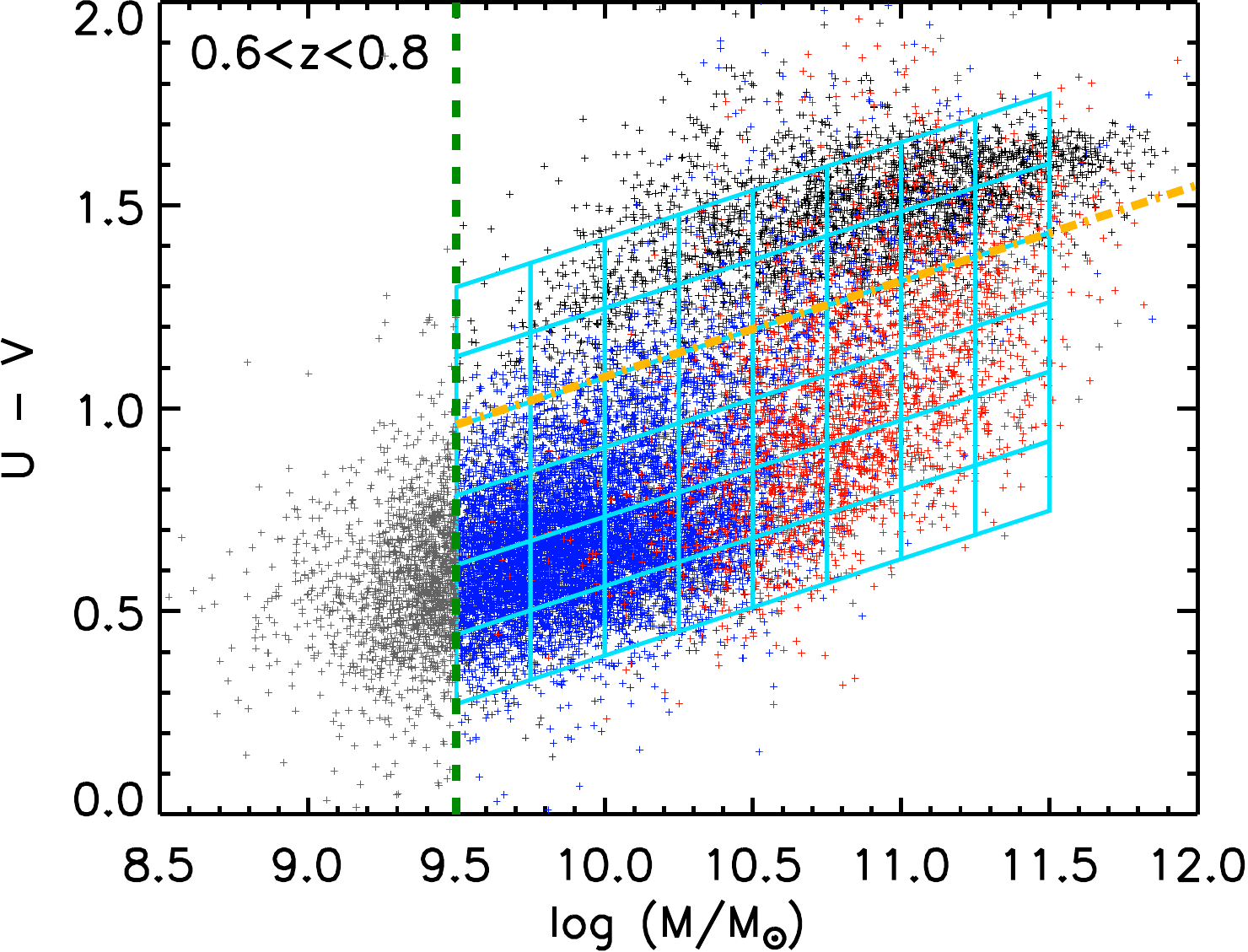} 
\caption{The diagram of mass versus rest-frame $U-V$ color for galaxies with $I_{F814W}<24$ and $0.6<z<0.8$ from COSMOS. The 24\,$\micron$-detected sources ({\it red crosses}) are mostly massive galaxies with $M>10^{10}M_\odot$. $24\,\micron$ undetected SFGs selected with $U-V-K$ selection (Figure~\ref{fig:uvk}) are shown as {\it blue crosses}. The {\it cyan} grid shows the color and mass cuts to split the 24\,$\micron$-undetected galaxies into subsamples for stacking. The division between red sequence and blue cloud described by Equation 1 is indicated by the {\it orange dash-dot} line. The {\it green-dashed} line marks the completeness limits for the magnitude cut $I<24$ \citep{Ilbert10}. Our mass cut of $\log(M/M_\odot)>9.5$ should be highly complete for SFGs which are dominated by the blue-cloud galaxies at the low-mass end. {\it Gray crosses} are galaxies with mass less than $10^{9.5}M_\odot$ [{\it See the online journal for a color version of this figure.}]}
\label{fig:masscolor}
\end{figure}

\section{Data and Sample Selection} \label{sec:data}

We use the multi-wavelength data set of COSMOS over an area of 1.48 square degrees, where  {\it HST}/ACS imaging \citep{Scoville07,Koekemoer07}, {\it Spitzer} Multi-band Imaging Photometer (MIPS) 24\,$\micron$ and Infrared Array Camera (IRAC) imaging \citep{Sanders07,LeFloch09}, and ground-based optical-to-near-Infrared (near-IR) imaging are available. In this work, we used the {\it HST}/ACS catalog \citep{Leauthaud07}, the  XMM-Newton X-ray Source Catalog \citep{Cappelluti09}, the multi-band photometric catalog \citep{Capak07}, and the photometric redshift catalogs \citep{Ilbert09,Salvato09}, all of which are publicly available.\footnote{http://irsa.ipac.caltech.edu/Missions/cosmos.html}

We start from sources that have secure {\it HST}/ACS detections (i.e., $I_{F814W}<24$). A matching radius of $0\farcs4$ is applied to cross-correlate the {\it HST}/ACS catalog with the multi-band photometric catalog. Nearest counterparts are selected. Active Galactic Nuclei (AGNs) in the XMM-Newton X-ray Source Catalog \citep{Cappelluti09} are removed. We use a radius of $1\farcs5$ in matching the multi-band catalog with our own $24\,\micron$ catalog, the construction of which will be described later in this section.

Stellar mass is derived from the rest-frame $K$-band luminosity following \citet{Arnouts07} with a Chabrier \citeyearpar{Chabrier03} Initial Mass Function (IMF). While rest-frame UV traces unobscured star-forming activity, $24\,\micron$ traces dust-obscured star formation. The {\it Spitzer}/MIPS $24\,\micron$ observations are about five times deeper in terms of SFR sensitivity than the {\it Herschel} coverage of the COSMOS field. Direct {\it Herschel} detections for a significant fraction of the SFG-population are hence only expected in the most massive galaxies \citep{Lutz11,Oliver12}. Therefore, for the present analysis we use only the deeper {\it Spitzer}/MIPS data.

\begin{figure}
\epsscale{1.08}
 \plotone{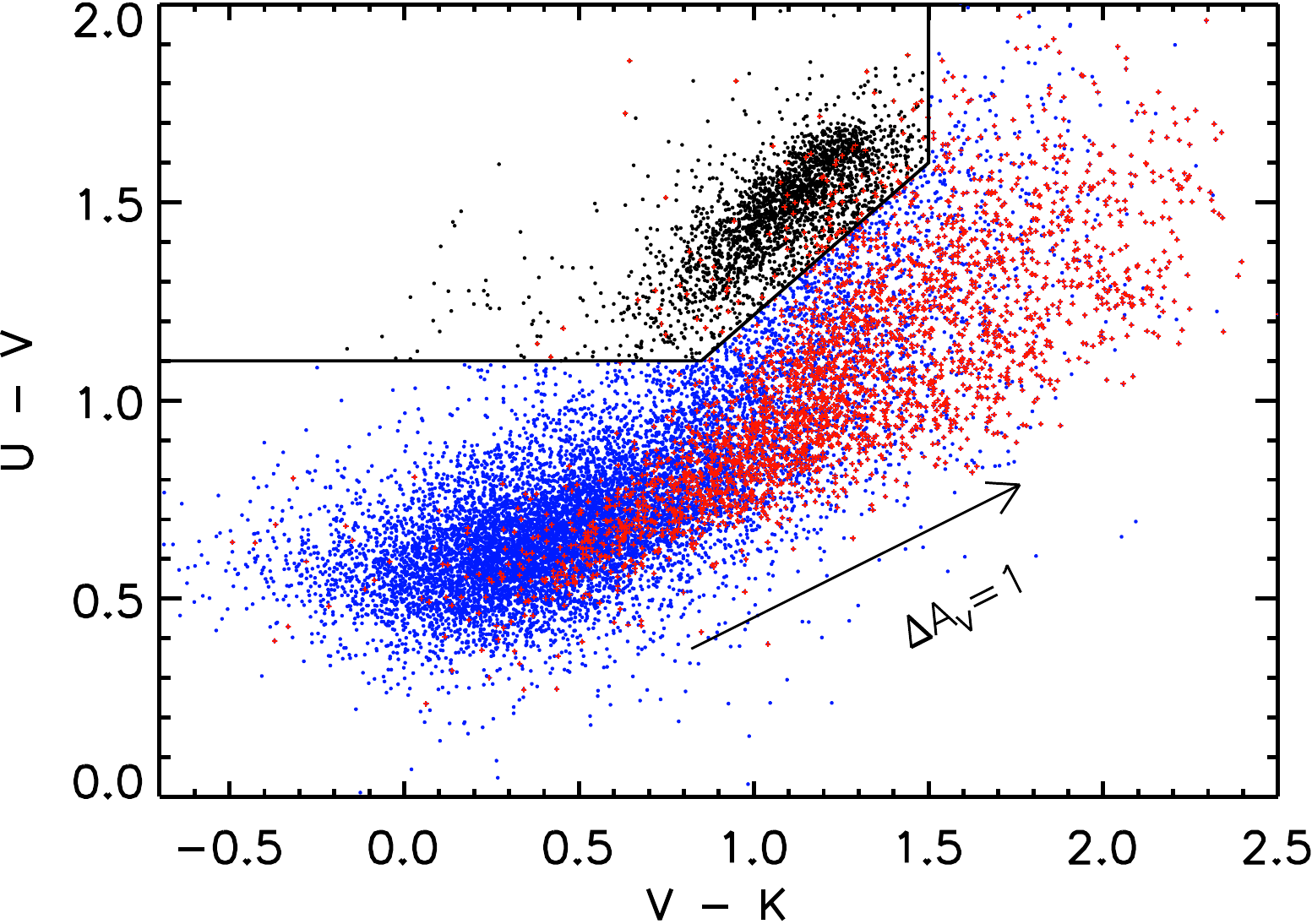}
\caption{The color-color diagram of 15\,598 galaxies with $\log (M/M_\odot)>9.5$ and $I_{F814W}<24$ between $0.6<z<0.8$ from COSMOS. The three lines ($V-K >1.5$, $U-V<$\,1.1+0.77\,$(V-K-0.85)$, and $U-V <1.1$) define the separation cuts between quiescent galaxies ({\it black dots} in the top-left region) and SFGs ({\it blue dots}). The {\it red crosses} are SFGs individually-detected at 24\,$\micron$. The arrow shows the effect of dust extinction with $A_{\rm V}=1$\,mag following the Calzetti Law. [{\it See the online journal for a color version of this figure.}]
}
\label{fig:uvk}
\end{figure}

We focus on galaxies in the redshift range of $0.6<z<0.8$ for the following reasons: 1) the cosmic SFR density at $z \sim 0.7$ is $\sim 5$ times higher than at the present-day and a large fraction of massive galaxies were still actively forming stars \citep[e.g.,][]{LK10}; 2) the sensitivity of current deep IR observation allows to individually detect low-mass galaxies at a low SFR level ($3\,M_\odot$\,yr$^{-1}$) at this epoch; 3) photometric redshifts are reliable because the most important spectral feature, the 4000\,\AA/Balmer break, is still within the optical window where the deepest and most well-sampled photometry is available; 4) the narrow redshift range $\Delta z=0.2$ reduces evolutionary effects on the SSFR distribution. A simple correction based on the linear correlation between redshift and SSFR is applied (see \S 4 for details).
After removing the X-ray sources, a total of 17\,294 galaxies are selected with $I_{F814W}<24$ and $0.6<z<0.8$ \citep[the photo-z uncertainty is $\sigma_{\Delta z/(1+z_{\rm s})}=0.012$ for $I<24$ and $z<1.25$,][]{Ilbert09}. Figure~\ref{fig:masscolor} shows the bimodal color-$M$ distribution of these galaxies (i.e., so-called {\it red sequence} and {\it blue cloud}) in the diagram of stellar mass versus rest-frame $U-V$ color. The red sequence can be described as 
\begin{equation} \label{equ:red}
 U-V > 0.96 + 0.24\,(\log \frac{M}{M_{\odot}} - 9.5)
\end{equation}
with a spread of 0.17\,dex ($1\sigma$).
Note that magnitude selection leads to a color-dependent completeness in stellar mass. Our selection criterion $I_{F814W}<24$ corresponds approximately to a completeness of $>70\%$ for galaxies with $M>10^{9.5} M_{\odot}$ although blue-cloud galaxies are more complete than red-sequence galaxies \citep{Ilbert10}. 

It is important to note that, heavily obscured AGNs which are usually X-ray faint may contribute to mid-IR emission and compromise the $24\,\micron$-derived SFR \citep{Fu10}. We use IRAC colors to identify obscured AGNs characterized by the ``mid-IR excess''. The IRAC catalogs are extracted from the deep IRAC science images by ourselves based on aperture photometry after subtracting surrounding sources with the best-fit Point Spread Function (PSF) (Zheng et al., in prep). The $5\sigma$ detection limits are 1.43, 2.45, 17.17 and 23.15\,$\mu$Jy for IRAC 3.6, 4.5, 5.6 and 8.0\,$\micron$, respectively. Of the 17\,294 galaxies, 12\,171 have counterparts in both IRAC 3.6 and 4.5\,$\micron$ catalogs; galaxies with $M>10^{9.8} M_{\odot}$ are nearly complete in the 3.6 \& 4.5\,$\micron$ matched catalogs; only 2\,252 have counterparts in all four IRAC bands. Of the 2\,252 objects, 29 are identified as obscured AGNs by the IRAC color selection given in \citet{Stern05}, and 10 with $f_{\rm 36}/f_{\rm 45} >1.1$ are also classified as obscured AGNs. 
Removing these 39 AGNs, there remain 15\,598 galaxies with $M> 10^{9.5} M_{\odot}$ for further analysis. 

We extracted 24\,$\micron$ catalog using a PSF-fitting method of \citet{Zheng06}. Of 15\,598 sample galaxies, 2\,758 are found to be 24\,$\micron$ sources with $f_{24}>55$\,$\mu$Jy ($3\sigma$) via cross-matching the two catalogs with a matching radius of $1\farcs5$. The 24\,$\micron$ depth allows us to detect obscured star formation at a rate of $>3$\,$M_{\odot}$\,yr$^{-1}$ for SFGs at $z=0.7$. 
As shown in Figure~\ref{fig:masscolor}, the 24\,$\micron$-detected sources mostly have $M>10^{10} M_{\odot}$. And a significant fraction of the 24\,$\micron$-detected objects fall on the red sequence. These are dusty red SFGs. 
We use the $U-V$ versus $V-K$ selection adapted from \citet{Williams09} to select SFGs, as shown in Figure~\ref{fig:uvk} for the 15\,598 galaxies with $M>10^{9.5} M_{\odot}$. The bulk of dusty red SFGs detected at 24\,$\micron$ are clearly separated from the quiescent galaxies. We count all 24\,$\micron$-detected galaxies as SFGs. Finally, we obtain a sample of 12\,614 SFGs with $M> 10^{9.5} M_{\odot}$, shown as blue and red symbols in Figures~\ref{fig:masscolor} and~\ref{fig:uvk}. It is clear that our sample is dominated by blue-cloud galaxies defined by $U-V \leq 0.96+0.24\,(\log \frac{M}{M_{\odot}} -9.5)$ at the low-mass end, and a significant fraction of high-mass SFGs are dusty ones with colors similar to the red-sequence galaxies. The 24\,$\micron$-undetected galaxies in the red sequence are mostly quiescent galaxies with little or no star formation. 
We consider our selection of SFGs complete above our mass limit.

\section{SFR Measurement}\label{sec:sfr}

\begin{figure*}
\plotone{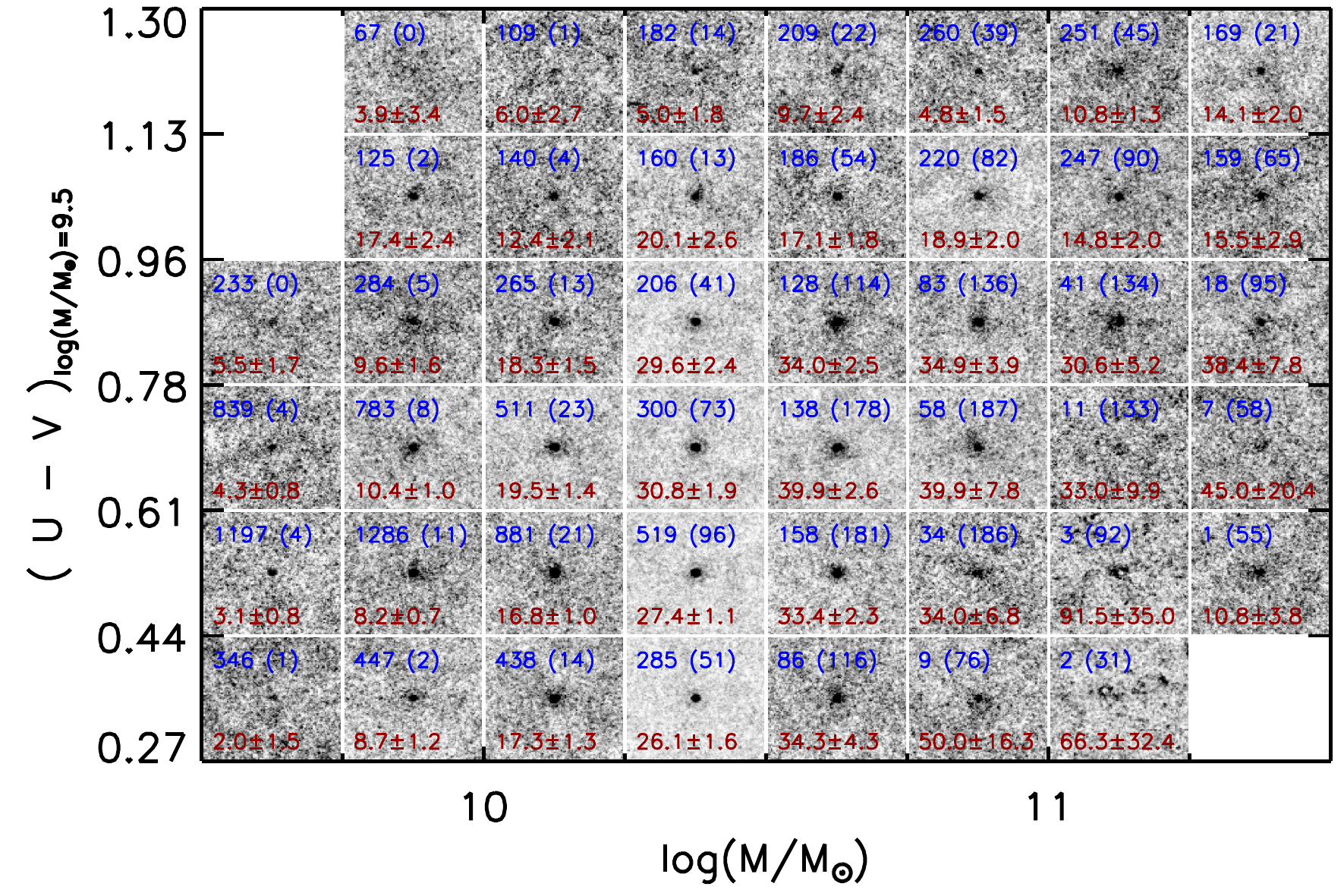}
\caption{The 24\,$\micron$ stacked stamp images for undetected SFGs in various color and mass bins. Colors are normalized to that of galaxies with $\log(M/M_\odot)=9.5$ with slopes and offsets corresponding to blocks in color-mass diagram (Figure~\ref{fig:masscolor}). Highly significant signals are detected in nearly all bins. The number of $24\,\micron$-undetected galaxies for stacking is labeled ({\it blue}) together with that of those detected
(in parentheses) in each stamp, where the mean stacking flux as well as its bootstrapping error are also labeled in {\it red}. [{\it See the online journal for a color version of this figure.}]}
\label{fig:stack24}
\end{figure*}

\subsection{UV and IR Luminosities} \label{subsec:uvlum}
We derive SFR from UV+IR luminosity. The same SFR estimator is applied to all SFGs. 
We measured the UV (1216$-$3000\,\AA) luminosity by integrating observed broad-band fluxes from the $NUV$ to the $V$-band. For low-mass SFGs that are too faint to be detected in the $NUV$ band, we fit their photometry between the $U$ and $V$-band with a library of stellar population synthesis models from \citet{BC03} for a range of ages with an $e$-folding star formation history. For simplicity, we assumed an e-folding time ($\tau$) of 1 Gyr for all of our models. The interpolated rest-frame UV luminosities are insensitive to the detailed model parameters.
A typical error of $L_{\rm UV}$ is $\sim30\%$, mostly determined by photometric errors in $U$ and $B$ band. 

As one can see from Figure~\ref{fig:masscolor}, most low-mass SFGs are undetected at 24\,$\micron$. Stacking technique is thus used to estimate the average 24\,$\micron$ flux for 24\,$\micron$-undetected galaxies. As shown in Figure~\ref{fig:masscolor} we divide the 24\,$\micron$-undetected galaxies into 6 color bins from $U-V$=\,0.27 to $U-V$=\,1.30 at $\log (M/M_{\odot})$=\,9.5 with $\Delta({U-V})$=\,0.17 and 8 mass bins from $\log (M/M_{\odot})$=\,9.5 to 11.5 with $\Delta \log M$=\,0.25 parallel to the red sequence described by Equation~\ref{equ:red}. 
Following \citet{Zheng06}, we carry out a 24\,$\micron$ mean stacking analysis for these subsamples and perform photometry within an aperture of 10 pixels ($\sim2$ FWHM). We cut image stamps from the PSF-subtracted 24\,$\micron$ mosaic centered at the positions of the galaxies in each bin and co-add them to obtain the stacked images. Figure~\ref{fig:stack24} shows the stacked images. The number of 24\,$\micron$-undetected galaxies used in stacking is indicated together with that of those detected (in parentheses) in each bin, where the mean-stacked flux and its bootstrapping error are also labeled in red. Clearly, secure signals are detected in nearly all stacked bins. The background noise of a bin of stacked images is inversely proportional to the square of the number of images stacked \citep{Zheng06}. The stacked 24\,$\micron$ fluxes reach a minimum of $\sim15$\,$\mu$Jy, i.e., significantly lower than the 3$\sigma$ detection limit (55\,$\mu$Jy).

In contrast to the case for $z>1.5$ where 24\,$\micron$ observation probes the rest-frame $<10\micron$ and far-IR observations (e.g., from {\it Herschel}) become critical to estimating SFR,  the 24\,$\micron$ observation essentially measures the rest-frame 13$-$15\,$\micron$ luminosity for $z\sim0.7$ SFGs, which is well correlated with the total luminosity \citep[e.g.][]{Rieke09}. 
We convert 24\,$\micron$ luminosity into the total IR (8$-$1000\,$\micron$) luminosity using a library of luminosity-dependent IR SED templates with an uncertainty of 0.13 dex from \citet{Rieke09}. Accounting for the plausibility that distant luminous IR galaxies (LIRGs; ${\rm SFR}>10$\,$M_\odot$\,yr$^{-1}$) are colder than those in the local universe \citep[e.g.][]{Zheng07b,Wuyts11a}, we adopt IR SEDs with their corresponding IR luminosities one order of magnitude lower for the LIRGs in our sample.\footnote{Adoption of the IR template from \citet{Wuyts08} gives consistent IR luminosity measurements within 0.15\,dex. } We calculate SFR following \citet{Bell05}:
\begin{equation} \label{equ:sfr}
{\rm SFR}= 9.8\times 10^{-11}\,(L_{\rm IR}+2.2\,L_{\rm UV}),
\end{equation}
where UV and IR luminosities are given in units of solar luminosity, and SFR is in  $M_\odot$\,yr$^{-1}$. The errors of SFR come from those in $L_{\rm IR}$ and $L_{\rm UV}$, where the former is affected by the measurement and conversion error of $L_{\rm24}$, as mentioned above. We have not taken into account the systematic error of the $24\,\micron$-derived $L_{\rm IR}$. We will discuss its potential impact in \S 4.

\subsection{Modeling  the distribution of 24\,$\micron$ fluxes for  individually-undetected galaxies} \label{subsec:mod}

Our goal is to examine the distribution of SSFRs as a function of galaxy stellar mass in an unbiased way. The stacking method, however, only gives the averaged 24\,$\micron$ flux for a subsample of 24\,$\micron$-undetected galaxies. The scatter of 24\,$\micron$ flux among these galaxies remains unknown. The adoption of the averaged 24\,$\micron$ flux for individual galaxies of the subsample would possibly underestimate the scatter in SSFR, particularly for low-mass SFGs.
 
\begin{figure*} 
\plottwo{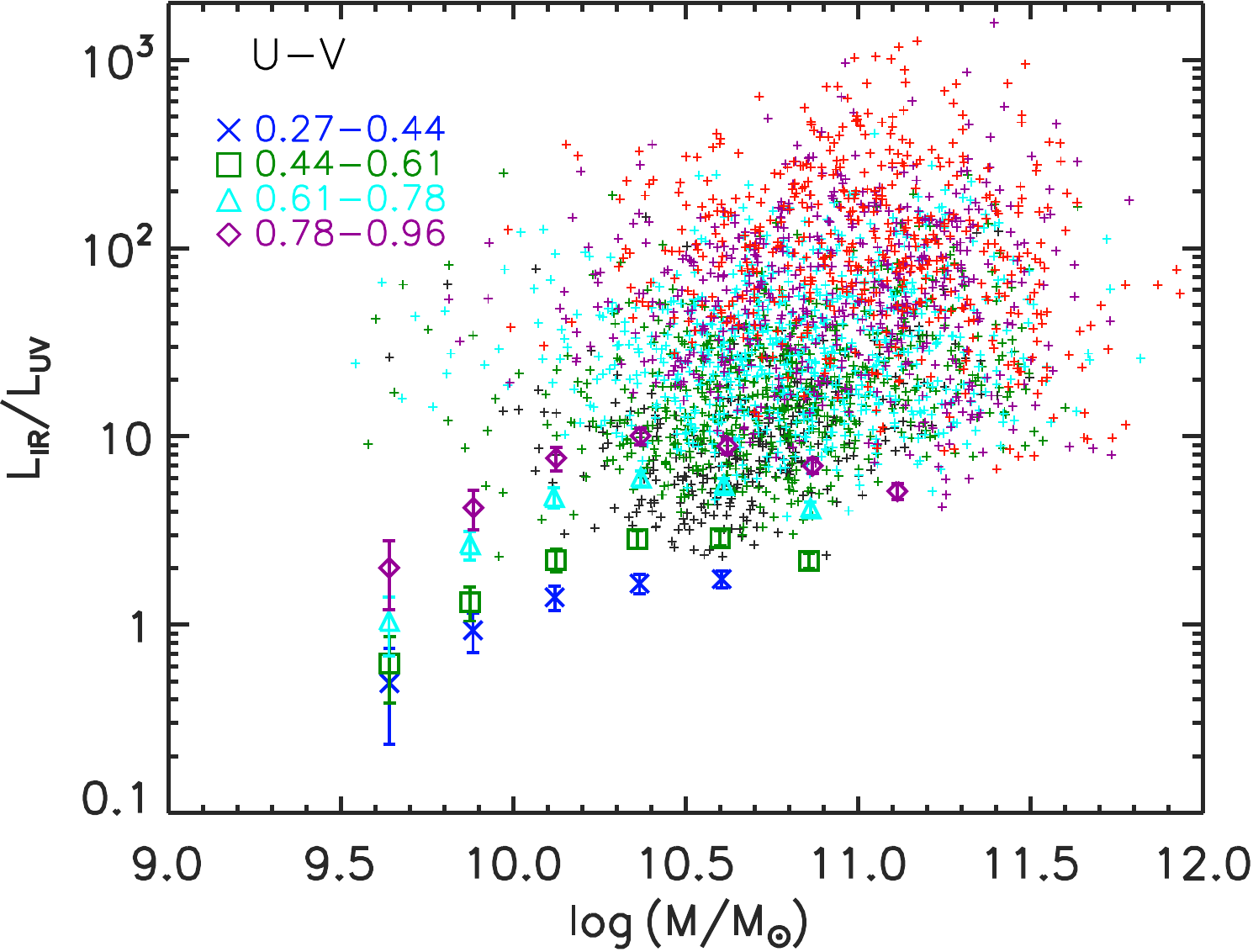}{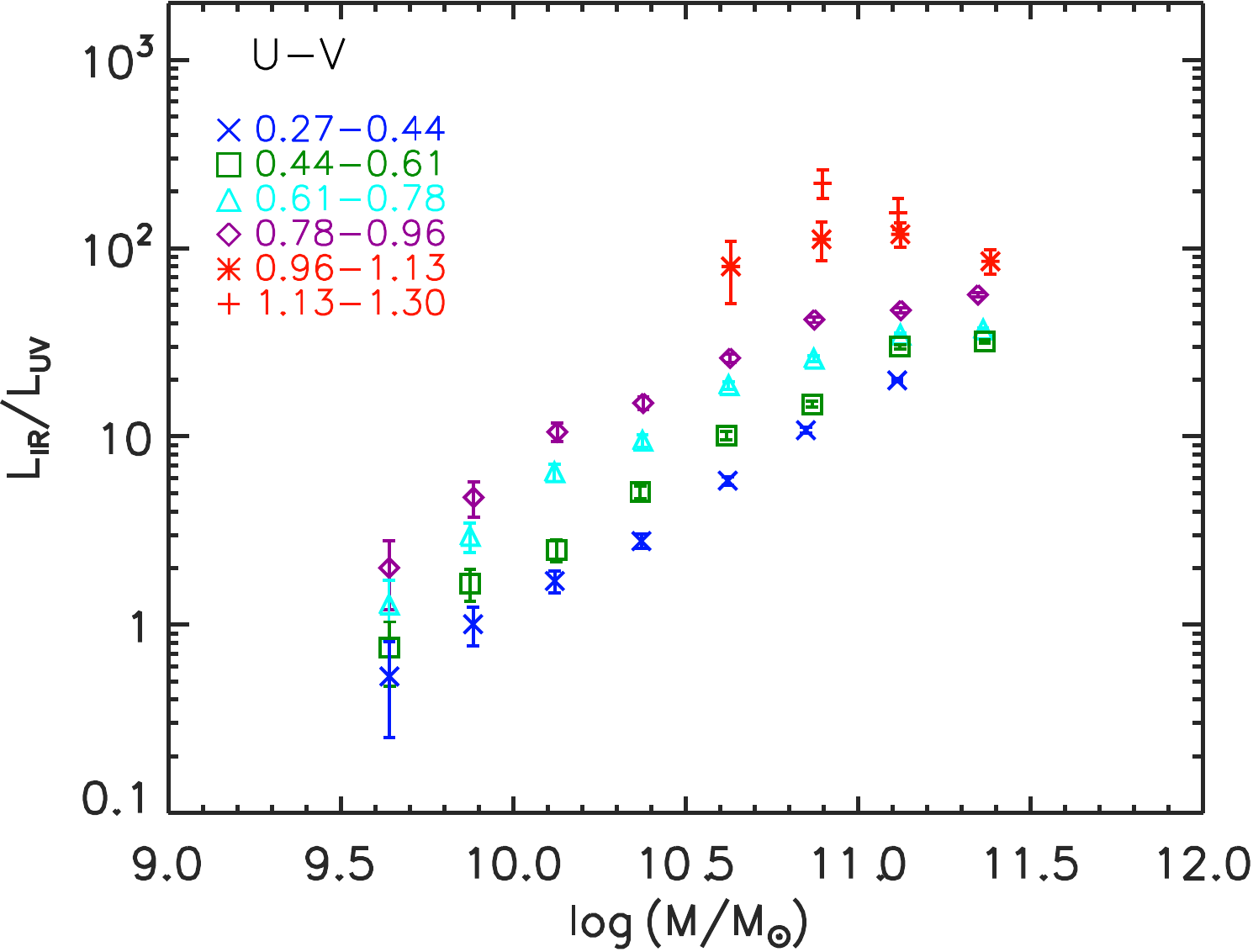}
\caption{{\it Left:} L$_{\rm IR}$/L$_{\rm UV}$ versus stellar mass for 24\,$\micron$-detected SFGs ({\it color} coded with $U-V$ color) and for subsamples of 24\,$\micron$-undetected SFGs divided by color and mass ({\it large symbols}). Stacking is used to derive the averaged 24\,$\micron$ fluxes for the latter. {\it Right:} the average $L_{\rm IR}/L_{\rm UV}$ as functions of stellar mass and $U-V$ color. Both 24\,$\micron$-detected and undetected SFGs are included. {\it Red symbols} are dusty red SFGs detected by $24\,\micron$. Extinction increases with stellar mass and color, with a large spread for $24\,\micron$ detected galaxies. [{\it See the online journal for a color version of this figure.}]
}
\label{fig:iruv}
\end{figure*}

Figure~\ref{fig:iruv} shows the extinction (i.e., $L_{\rm IR}/L_{\rm UV}$) as functions of stellar mass and $U-V$ color for our sample SFGs. The left panel shows the large spread of $L_{\rm IR}/L_{\rm UV}$ for individual 24\,$\micron$-detected SFGs and the average results for 24\,$\micron$-undetected SFGs split into mass and color bins. The right panel gives the average $L_{\rm IR}/L_{\rm UV}$ for given stellar mass and $U-V$ color bins.\footnote{The $\frac{\sum L_{\rm IR}}{\sum L_{\rm UV}}$ is utilized as the average value for a subsample.  Such is more meaningful than $\overline{{L_{\rm IR}}/{L_{\rm UV}}}$. Details about the differences between these two algorithms could be found in \citet{Brinchmann04}.}
It is clear from Figure~\ref{fig:iruv} that $L_{\rm IR}/L_{\rm UV}$ generally increases with stellar mass and $U-V$ color, while the scatter in $L_{\rm IR}/L_{\rm UV}$ is rather large, i.e., approximately $1\sigma\sim0.6$\,dex for high-mass SFGs \citep[see also][for $z\sim 2$ results]{Pannella09,Whitaker12,Heinis13}.

\begin{figure} 
\plotone{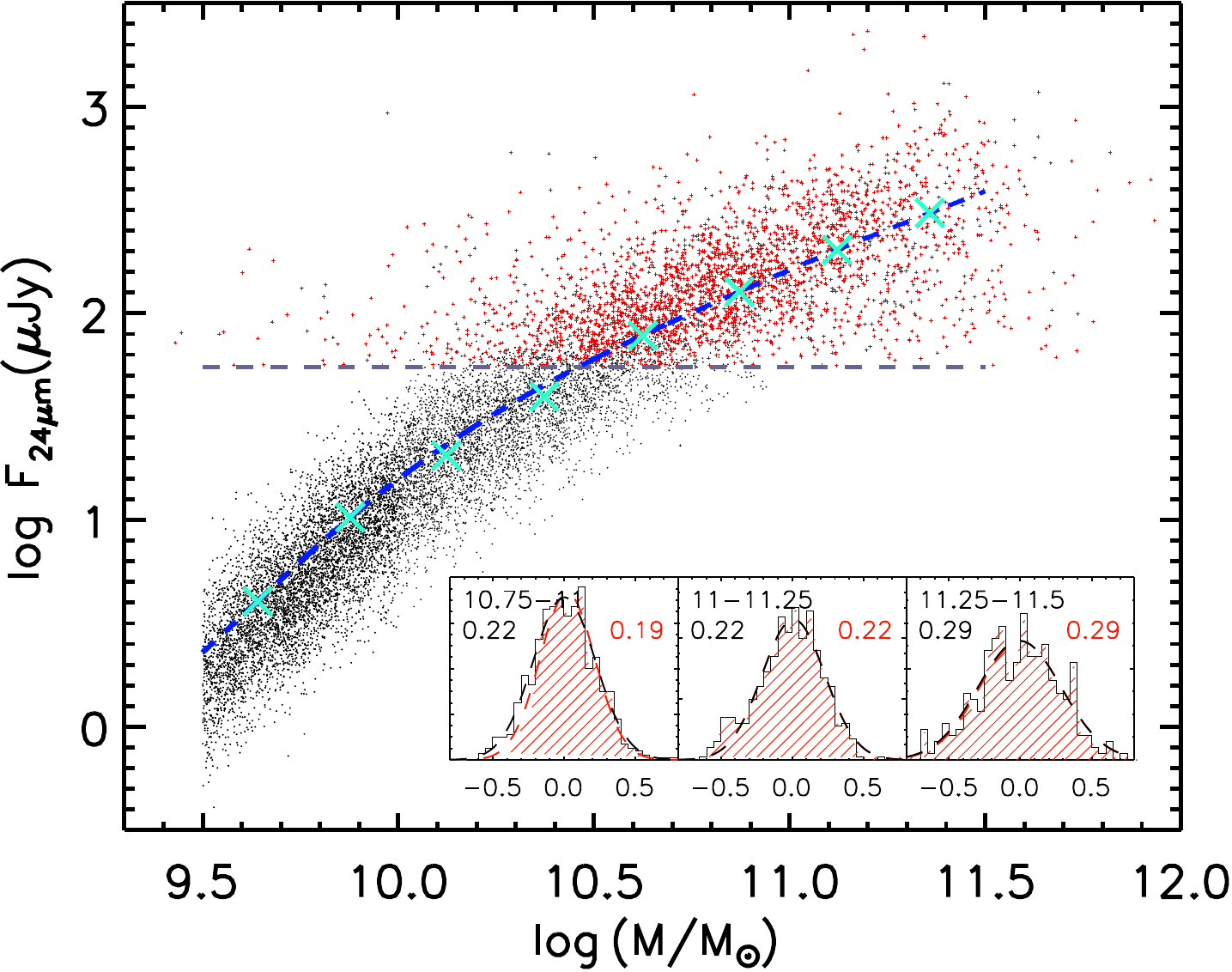}
\caption{The modeled 24\,$\micron$ fluxes ({\it black} dots) assuming a constant scatter of 0.22\,dex and the mean stacked flux density ({\it cyan X}) derived through stacking for 24\,$\micron$-undetected sample SFGs are plotted together with 24\,$\micron$-detected SFGs ({\it red} crosses). The {\it blue dashed} line is the best fit to the averaged 24\,$\micron$ fluxes over all SFGs in each mass bin from $\log (M/M_\odot)=9.5$ to 11.5. The horizontal dashed-thick line is the 3$\sigma$ (55\,$\mu$Jy) detection limit.  The inner panels show the normalized distributions of 24\,$\micron$ flux densities in three high mass bins, with the mass ranges labeled at the top. The scatters of 24\,$\micron$ fluxes are also marked (left: all SFGs; right: 24\,$\micron$-detected SFGs). The hatched histogram represents 24\,$\micron$-detected SFGs while the un-hatched histogram includes 24\,$\micron$-undetected SFGs. Here we introduce the scatter in $24\,\micron$ flux to estimate its impact to the derived SFRs. [{\it See the online journal for a color version of this figure.}]}
\label{fig:f24}
\end{figure}

Local SFGs with lower SFR tend to have $L_{\rm IR}/L_{\rm UV}$ ratios that are lower and have a smaller spread \citep{Bothwell11}. This implies that low-mass SFGs are unlikely to show a larger spread in IR luminosity than high-mass SFGs given that the low-mass SFGs are dominated by blue-cloud galaxies with similar $U-V$ color. The distribution of $\log (f_{24}/M)$ tends to be broader with increasing stellar mass: 1$\sigma$ scatter is 0.19, 0.22 and 0.29 for $\log (M/M_\odot)$=10.75$-$11, 11$-$11.25 and 11.25$-$11.5, respectively (see the inner panel of Figure~\ref{fig:f24}). The spread of $\log f_{24}$ at the high-mass end can thus provide an upper limit for that of SFGs at lower masses. However, even if an unrealistically large spread of 0.5\,dex were applied to low-mass bins, our results would not change significantly.
Nevertheless, taking $1\sigma=0.22$ as the spread of $\log f_{24}$, together with the mean of 24\,$\micron$ fluxes determined from stacking, we can assign 24\,$\micron$ fluxes for the 24\,$\micron$-undetected SFGs. 
A cubic spline function is used to fit the mean 24\,$\micron$ fluxes and the interpolated values from the best-fit function are used in distributing 24\,$\micron$ fluxes. Figure~\ref{fig:f24} presents the distribution of the modeled 24\,$\micron$ fluxes as a function of galaxy mass, together with individually-detected 24\,$\micron$ fluxes. 
We notice that there are 39 SFGs with stellar mass $<10^{10}\,M_\odot$ having $24\,\micron$ fluxes above the detection limit, and far off the distribution of $24\,\micron$-undetected SFGs at low mass. By examining their {\it HST}/ACS F814W images, we find these $24\,\micron$-detected low-mass SFGs are mostly companioned by a nearby bright source and only a small fraction (12 of 39) appear to be isolated within the matching radius of $1\farcs5$. This hints that the $24\,\micron$ fluxes may be substantially overestimated for the majority of the 39 low-mass SFGs  due to the contamination from nearby sources. We ignore these data points in modeling $\log f_{24}$ for low-mass SFGs. This has little effect on the scatter of $\log f_{24}$.

\begin{figure} 
   \plotone{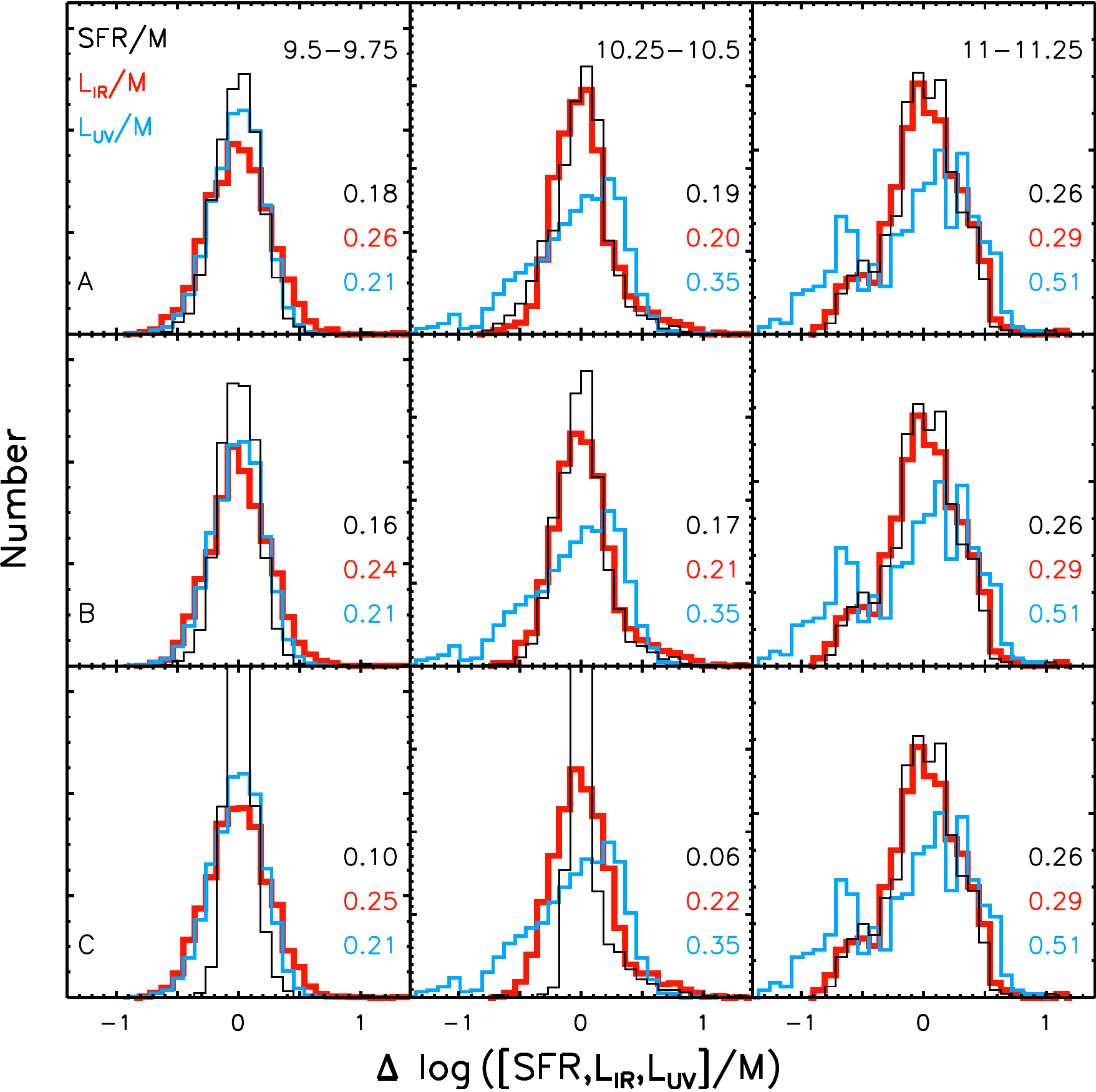}
\caption{The distributions of ${\rm SFR}/M$ ({\it black-thin}), $L_{\rm IR}/M$ ({\it red-thick}) and $L_{\rm UV}/M$ ({\it blue}) for SFGs with $\log (M/M_\odot)=9.5-$9.75 ({\it left}), 10.25$-$10.5 ({\it middle}) and 11$-$11.25 ({\it right}). The {\it top} panels show the results from Model {\bf A}: the modeled 24\,$\micron$ fluxes are sorted and assigned to the 24\,$\micron$-undetected SFGs sorted by the predicted IR luminosities (and predicted 24\,$\micron$ fluxes) from the UV luminosities using the relations in Figure~\ref{fig:iruv} in combination with stellar mass and $U-V$ color; 
The {\it middle} panels refer to the model {\bf B}: the modeled $f_{24}$ fluxes are randomly assigned to 24\,$\micron$-undetected SFGs of similar masses. The {\it bottom} panels refer to the model {\bf C}: the modeled $f_{24}$ fluxes of 24\,$\micron$-undetected SFGs are assigned according to the UV slope. Scatters of the best-fit Gaussian profiles in each panel are listed. We measure similar scatters for models A and B. In model C, the distributions are systematically narrower because UV slope is generally correlated with extinction.
[{\it See the online journal for a color version of this figure.}]}
\label{fig:simu}
\end{figure}

We assign the modeled 24\,$\micron$ fluxes to 24\,$\micron$-undetected galaxies on a model-dependent basis.
Three cases are considered: case A --- the modeled 24\,$\micron$ fluxes are sorted and assigned to the 24\,$\micron$-undetected SFGs sorted by the predicted IR luminosities (and predicted 24\,$\micron$ fluxes) from the UV luminosities using the interpolated  $L_{\rm IR}/L_{\rm UV}$ from stellar mass and $U-V$ color (Figure~\ref{fig:iruv}); 
case B --- the modeled 24\,$\micron$ fluxes are randomly assigned to the 24\,$\micron$-undetected SFGs; case C --- the modeled 24\,$\micron$ fluxes are assigned to 24\,$\micron$-undetected SFGs sorted by the slope of the UV continuum defined by \citet[][]{Calzetti94}. 
We then use these assigned 24\,$\micron$ fluxes to calculate the IR luminosities and SFRs for each SFG.
Figure~\ref{fig:simu} presents the distributions of $L_{\rm IR}, L_{\rm UV}$, and SFR  in three stellar mass bins: $\log (M/M_\odot)$=9.5$-$9.75, 10.25$-$10.5 and 11$-$11.25.  The three parameters are normalized to galaxy stellar mass. Panels from the top to the bottom refer to case A to C. The scatters of these distributions are estimated from the best-fit Gaussian functions and listed to the right of each panel.
We find that for $\log (M/M_\odot)$=9.5$-$9.75 the modeled $L_{\rm IR}$ is comparable to the observed $L_{\rm UV}$ in terms of the scatter and intensity (see Figure~\ref{fig:iruv} for $L_{\rm IR}/L_{\rm UV} \sim1-2$), and the scatter of the corresponding SFR  is smaller than that of $L_{\rm IR}$ or $L_{\rm UV}$ in case A and the scatter is even smaller in case B when the two luminosities are uncorrelated. 
This is not surprising because the sum of two sets of random numbers that satisfy the same log-normal distribution follows a log-normal distribution with a smaller scatter.
Note that $L_{\rm UV}$ contributes more to SFR than $L_{\rm IR}$ at $L_{\rm IR}/L_{\rm UV} <\sim2$ (see Eq.~\ref{equ:sfr}). For the other two mass bins, $L_{\rm UV}$ spreads broader than $L_{\rm IR}$ but represents only a small fraction of SFR (as seen in Figure~\ref{fig:iruv}); and SFR thus roughly follows $L_{\rm IR}$ in scatter. 
 Given that the combination of IR and UV luminosities is a measure of SFR and dust extinction adds additional dispersions to the luminosities,  the scatter of either UV or IR luminosity is expected to be larger than the scatter of SFR.
It is worth noting that the scatter in SSFR of case C is much smaller, compared to the scatters for case A and B. This is expected because the UV slope is generally correlated with the extinction and therefore the combination of UV and IR luminosities in case C substantially suppresses the scatter in extinction.
We conclude that: 1) SFR follows $L_{\rm IR}$ in scatter for massive SFGs and $L_{\rm UV}$ gradually controls the scatter of SFR towards the low-mass end; and 2) the scatter of 0.22\,dex in $f_{24}$ for 24\,$\micron$-undetected SFGs provides a reasonable upper limit to constrain the scatter of SFR. 
We emphasize that even with an unrealistically large scatter (e.g., 0.5\,dex) our results would not change because in the low-mass bins the scatter in SFR is dominated by that of the UV luminosity.

We consider case A, in which $f_{24}$ are assigned following the interpolated relation of $L_{\rm IR}/L_{\rm UV}$ with stellar mass and optical color, to be the most realistic scenario and adopt it for the rest of the analysis. By using this approach to assign modeled fluxes $f_{24}$ to all 24\,$\micron$-undetected SFGs, we recover the whole range of the $L_{\rm IR}/L_{\rm UV}$ ratios and obtain an SFR estimate for every SFG in our sample.

\section{The  slope and scatter of the main sequence}\label{sec:ssfr}

\begin{figure} 
\epsscale{1.1}
\plotone{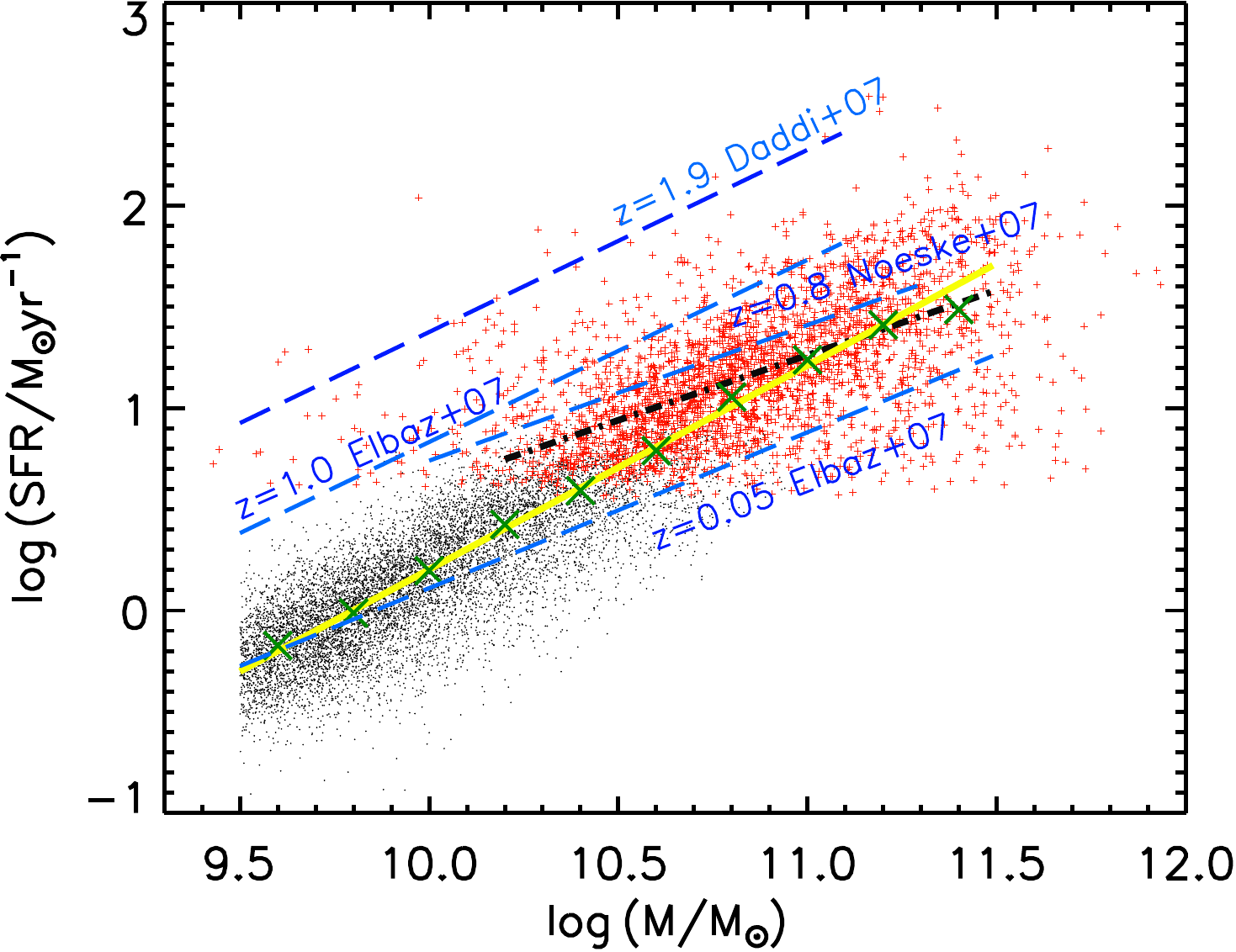}
\caption{The mass-SFR relation of our sample SFGs ({\it red crosses:} $24\,\micron$-detected, {\it black dots:} $24\,\micron$-undetected). The {\it yellow solid} line is the best power-law fit to the relation, suggesting a slope of 1.01. The best-fit to the mean SFRs ({\it green X}) also gives the same slope within uncertainties. The mass-SFR relations from previous works are also shown for comparison \citep{Noeske07a,Elbaz07,Daddi07a}. The {\it black dash dot} line with a slope of 0.64 is the best fit to the 24\,$\micron$-detected SFGs. 
\citet{Noeske07a} gave a shallower slope at the same redshift because of
incomplete SFG sample selection. In our case, the slope of the mass-SFR
relation is mostly determined by the dominant population of $24\,\micron$-undetected SFGs.
[{\it See the online journal for a color version of this figure.}]}
\label{fig:masssfr}
\end{figure}

The cosmic SFR density declines rapidly since $z\sim2$ \citep[e.g.,][]{Karim11}.
The strong evolution may broaden the mass-SFR relation for SFGs over a large redshift range. For our SFG sample at $0.6<z<0.8$, the cosmic SFR density decreases about 50\% from $z=0.8$ to $z=0.6$.  We thus scale the SFRs to $z=0.7$ and correct the evolution assuming $
{\rm SFR}(z)\propto (1+z)^{3.4} $
from \citet{LK10}. 
Figure~\ref{fig:masssfr} shows the SFR as a function of stellar mass for our sample. Fitting the relation with a straight line, we obtain 
\begin{equation}
\log ({\rm SFR}/M_\odot\,{\rm yr}^{-1}) = \alpha \log (M/M_\odot) + \beta
\end{equation}
with $\alpha=1.01\pm0.01$ and $\beta=-9.88\pm0.05$, shown as the yellow solid line in Figure~\ref{fig:masssfr}. The slope of unity suggests a constant SSFR for SFGs over the entire mass range examined here. The relations given in previous works are also plotted for comparison. The slope obtained is consistent with the results given in  \citet{Elbaz07} for $z\sim1$ SFGs, \citet{Daddi07a} for $z\sim1.9$ SFGs and \citet{Karim11} over $0.2<z<3$. We stress that sample selection is critical for determining the slope of the mass-SFR relation. For instance, fitting only to the 24\,$\micron$-detected galaxies would yield a much shallower slope, as shown in Figure~\ref{fig:masssfr}.  The slope of the relation is largely determined by the mean SFRs at given masses. The stacking analysis for low-mass SFGs is thus crucial to determine the slope. 

\begin{figure*} 
\plotone{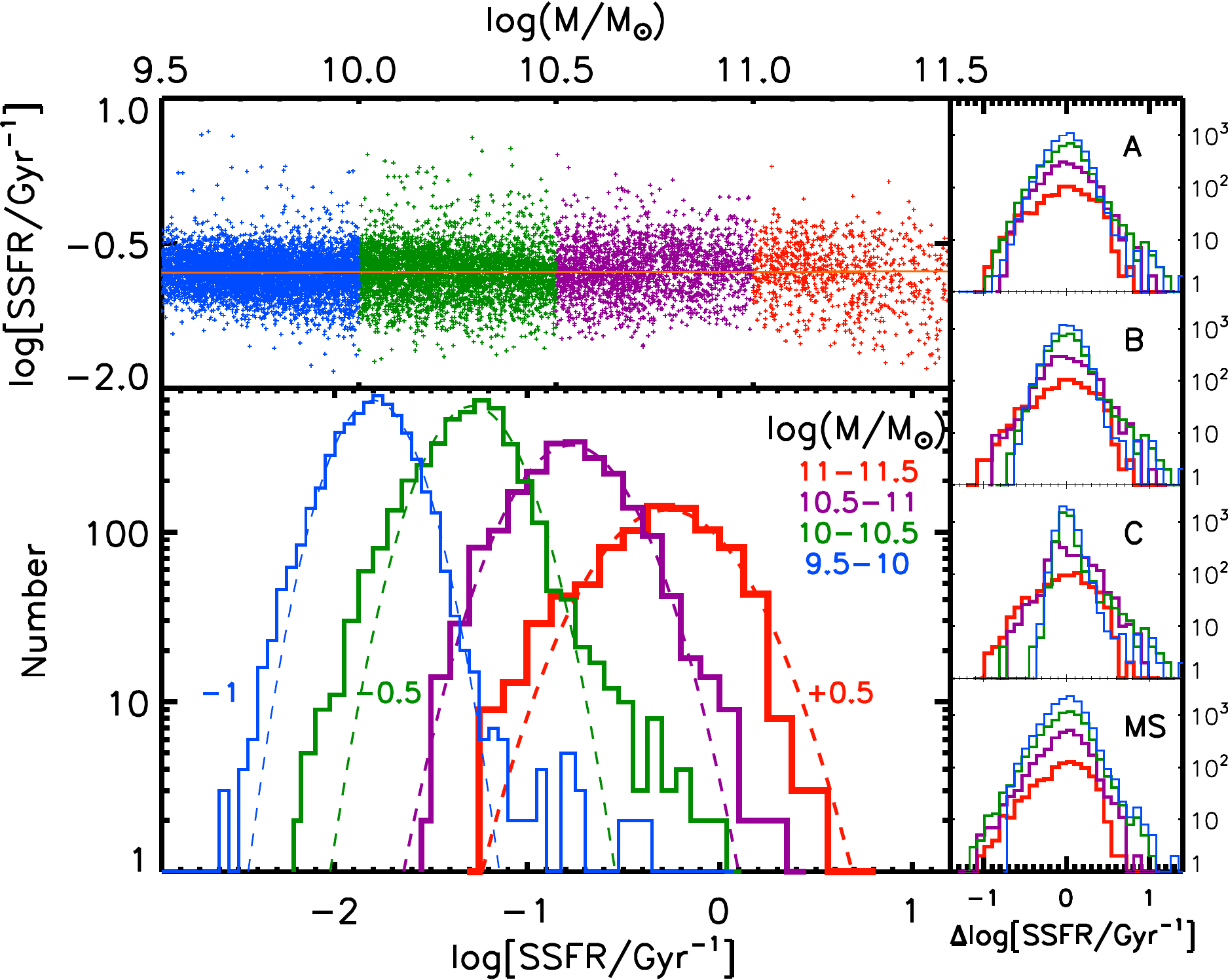}
\caption{The intrinsic scatter of the main-sequence increases with stellar mass, indicating more violent star formation in more massive galaxies.
The {\it top} panel: SSFR versus stellar mass for our sample SFGs with the best-fit relation shown as the orange solid line. The {\it bottom} panel: the distributions of SSFR for SFGs in four mass bins from $\log (M/M_\odot)$=9.5 to 11.5.  The distributions are shifted by $-$1, $-$0.5, 0, $+$0.5 for clarity. The {\it dashed} curves show the best-fit Gaussian functions to the distributions. The {\it right} panels: the SSFR distributions of the four mass bins. The distribution curves are aligned with each other by matching peak positions. The first three panels from the top refer to the models ``case A to C''  for assigning $24\,\micron$ fluxes to the $24\,\micron$-undetected SFGs; the panel labeled by ``MS'' is the same as the panel``A'' but for SSFR distributions based on the Elbaz main-sequence SED \citep{Elbaz11}. Regardless of the model, the scatter in SSFR increases with stellar mass. [{\it See the online journal for a color version of this figure.}]}
\label{fig:ssfr}
\end{figure*}

We now examine the scatter of the SSFR as a function of stellar mass.  Figure~\ref{fig:ssfr} presents. The top panel shows the SSFR versus stellar mass, and the bottom panel shows the distributions of SSFR in logarithm for four mass bins from $\log (M/M_\odot)$=\,9.5 to 11.5, with the median SSFR being 0.16, 0.16, 0.17, 0.16\,Gyr$^{-1}$, respectively. The evolution of SSFR for all models is shown in the right panels of Figure~\ref{fig:ssfr}. Since the median of SSFR is almost constant at $\log ({\rm SSFR}/{\rm Gyr}^{-1}) \sim -0.8$ over the stellar mass range, in the bottom panel we shift the four distribution curves along the $x$-axis by $-$1, $-$0.5, 0, +0.5, respectively, for clarity. Fitting the distributions of $\log ({\rm SSFR})$ with a Gaussian function gives 1\,$\sigma$ scatter to be 0.18, 0.21, 0.26, 0.31\,dex for the four mass bins, respectively, with uncertainties less than 0.01\,dex. The Gaussian profiles fit the observed distributions really well. It is clear from our analysis that {\it the spread of SSFR increases with galaxy stellar mass}. Even if we mix the IR luminosity in each mass bin in a random way (i.e., Case B), dispersions of Gaussian distributions will be 0.16, 0.18, 0.25 and 0.31\,dex, leading to the same conclusion. SSFR estimated with IR luminosity determined from a crude $L_{\rm IR}-$UV slope correlation (Case C) has dispersions of 0.09, 0.08, 0.23 and 0.31\,dex. As mentioned in \S \ref{subsec:mod}, the scatters for the two low-mass bins, which are dominated by 24\,$\micron$-undetected SFGs, are possibly overestimated since the scatter of 0.22\,dex in $f_{24}/M$ from the high-mass end is adopted in modeling the $f_{24}$ distribution. Moreover,  the systematic uncertainties in converting 24\,$\micron$ fluxes into total IR luminosities tend to enforce the SSFR scatter-mass correlation since SFR is more controlled by IR luminosity in higher mass bins. 
Additionally, the evidence of broadening of the relation between IR-color and $L_{\rm IR}$\citep{Chapman03} also implies a larger dispersion of SSFR at higher stellar mass. However, we note that -- when the upper bound of the highest mass bin is reduced from $10^{11.5}$ to $10^{11.25} M_\odot$ -- the measured scatter drops from 0.31 to $\sim0.26$\,dex and the SSFR scatter-mass correlation thereby becomes slightly less significant. An artificial broadening of the SSFR-distributions at high mass might also arise as a result of the super-linear scaling between $L_{\rm MIR}$-to-$L_{\rm IR}$ ratios and increasing $L_{\rm IR}$ \citep{Nordon12}. For a simple test, a constant IR-color SED template, which is suggested for Main-Sequence SFGs \citep{Elbaz11}, is used to check the significance of the results of increasing scatter. Values of 0.19, 0.22, 0.21, 0.26\,dex (with uncertainties of $\sim 0.01$\,dex) are thus derived, indicating a mass-dependence of the scatter as well, albeit less significant.
We are expecting a better constraint of this trend from more detailed SED fitting with {\it Herschel} observation.

Furthermore, the SSFR distribution appears distinct above and below a stellar mass of $\log(M/M_\odot)$=10.5; a significant excess from the best-fit lognormal function exists at the high-SSFR end of the observed SSFR distribution for two low-mass bins in Figure~\ref{fig:ssfr}.  Such excesses are also reported at $z\sim2$ and the outliers off the mass-SFR relation are often attributed to starbursts with enhanced star formation \citep{Rodighiero11}. The enhancement of star formation is often linked to violent processes such as galaxy interactions/merging, compared to star formation in the `normal' mode driven by secular processes.  
Interestingly, it is the mass bin of $\log (M/M_\odot)$=10$-$10.5 which exhibits the most prominent excess of outliers, whereas the two high-mass bins show weak or no such excesses. It appears that this feature reflects a mixture of violent (e.g., major/minor mergers) and quiescent processes for regulating star formation in this mass regime at $z\sim0.7$.

On the other hand, a starburst often refers to a galaxy forming a large number of new stars relative to existing old stars on timescales much shorter than the Hubble time. Here we identify a galaxy to be a starburst with the criterion of ${\rm SSFR}>0.3$\,Gyr$^{-1}$, which means that the galaxy would increase 30\% of its stellar mass within 1\,Gyr. We note that our threshold of starbursts is different from that used to pick up outliers. However, ${\rm SSFR}=0.3$\,Gyr$^{-1}$ matches $\sim0.7\,\sigma$ of the main sequence for SFGs with $\log (M/M_\odot)$=10$-$10.5 well. For the two high-mass bins, where only a few outliers can be found, the selection cut ${\rm SSFR}>0.3$\,Gyr$^{-1}$ picks up SFGs at the high SFR end. The chosen cut is able to pick up the majority of outliers in each mass bin. 
Such starbursts account for 7, 10, 18 and 19 percents of the SFG population in the four mass bins from $\log(M/M_\odot)$=\,9.5 to 11.5, respectively, and are responsible for 19\%, 27\%, 40\% and 40\% of the total SFR in each mass bin. This indicates that the mechanisms igniting starbursts (e.g., interactions/merging) are not the major processes driving the evolution of SFGs. Instead, the bulk of growth through star formation is associated with smooth processes regulating the mass-SFR sequence and its scatter. 
\citet{Rodighiero11} estimated that starbursts selected by SSFR $>2.5\,\sigma$ of the main sequence contribute to $\sim$ 10\% of the SFR density at $z=2$. Adopting the same selection, we find these outliers contribute to 9\%, 10\%, 7\% and 2\% of the total SFR for the four mass bins from $10^{9.5}$ to $10^{11.5}\,M_\odot$, respectively. Therefore, the conclusion remains unchanged: only a small fraction of star formation is driven by violent starbursts at $z\sim0.7$.

\section{Discussion and Conclusion} \label{sec:discussion}

We use a sample of 12\,614 SFGs with $0.6<z<0.8$ and $M>10^{9.5}M_\odot$ from COSMOS to carry out a detailed analysis of the mass-SFR relation, aimed at investigating the intrinsic properties of the relation via reducing the observational uncertainties as much as possible. 
The SFG sample is selected with the $U-V-K$ method adapted from \citet{Williams09} (Figure~\ref{fig:uvk}), and supplemented with dusty red SFGs detected by {\it Spitzer} 24\,$\micron$ observations.
Using the public multi-wavelength data set in the COSMOS field, we estimate SFR from UV+IR luminosity. {\it Spitzer} 24\,$\micron$ fluxes are used to estimate IR luminosities based on the luminosity-dependent IR SED templates from \citet{Rieke09}. The UV luminosities are derived from the observed $NUV$-to-$V$-band SEDs. A stacking technique is utilized to derive the average 24\,$\micron$ fluxes for SFGs which are individually undetected at 24\,$\micron$. 
The fraction of the 24\,$\micron$-undetected SFGs is 99\%, 90\%, 33\% and 0\% for the four equally spaced mass bins from $10^{9.5}M_\odot$ to $10^{11.5}M_\odot$, respectively. Assuming that $f_{24}/M$ for low-mass SFGs also follows a log-normal distribution with the same scatter as that for high-mass SFGs, we model 24\,$\micron$ fluxes for the 24\,$\micron$-undetected SFGs in a statistic way. After that, three methods are tested in assigning the modeled 24\,$\micron$ fluxes to the 24\,$\micron$-undetected SFGs. From the test, we consider the method (``case A'') in which 24\,$\micron$ fluxes are assigned following the interpolated relation of $L_{\rm IR}/L_{\rm UV}$ with stellar mass and optical color to be the most realistic scenario. This approach is adopted to recover the whole range of the $L_{\rm IR}/L_{\rm UV}$ ratios and obtain an SFR estimate for every SFG in our sample. We point out that SFR follows $L_{\rm IR}$ in scatter for high-mass SFGs and $L_{\rm UV}$ gradually regulates the scatter of SFR towards the low-mass end.

We confirm that the mass-SFR relation has a slope close to unity \citep[see also][]{Elbaz07,Daddi07a,Karim11,Gilbank11,Salmi12}. Such a slope can be obtained by fitting either the entire SFG sample dominated by low-mass ones, or the mean/median SFRs at given stellar masses. It is worth noting that the slope is sensitive to the sample selection. The bulk of low-mass SFGs are intrinsically faint in the IR. Missing such objects in an SFG sample would result in an overestimate of the average SFR for low-mass SFGs and a decrease of SSFR with increasing stellar mass. 
On the other hand, dusty SFGs with optical color similar to the red-sequence galaxies, which account for roughly one quarter of 24\,$\micron$-detected SFGs with $\log (M/M_\odot)>10.5$,  play a crucial role in shaping the high-end of the mass-SFR relation. The vast majority of such dusty SFGs can be successfully separated from the red-sequence galaxies in the $U-V-K$ diagram shown in Figure~\ref{fig:uvk}. 
A slope of unity for the mass-SFR relation means that SFGs of different stellar masses have a nearly constant SSFR in the population average sense. 
Together with the fact that the slope of the mass-SFR relation does not change much from $z\sim2$ to the present day \citep{Karim11}, this supports the picture that star formation in SFGs is generally driven by gas accretion from a gradually decreasing gas reservoir in galaxy halos from high to low redshift \citep[e.g.,][]{Dutton10,Behroozi10}. 

Interestingly, we find that the scatter along the mass-SFR relation is unlikely constant based on our partially model-based SFR estimation: for our prefered modeling scenario (``case A''), the 1\,$\sigma$ scatter in SSFR is 0.18, 0.21, 0.26 and 0.31 for the four equally spaced mass bins from 10$^{9.5}\,M_\odot$ to 10$^{11.5}\,M_\odot$, respectively. {\it This implies that the scatter of SSFRs increases with stellar mass}. We emphasize that this tendency is unlikely induced by observational effects. 
Figure~\ref{fig:iruv} shows that the extinction $L_{\rm IR}/L_{\rm UV}$ globally increases with galaxy stellar mass and color. 
The typical $L_{\rm IR}/L_{\rm UV}$ gradually decreases from a few $\times$10 for high-mass SFGs to $\sim$1 for low-mass SFGs although the dispersion is quite large (0.6\,dex). SFR is thus mainly traced by IR luminosity at the high-mass end and gradually by UV luminosity to the low-mass end, as is the SFR scatter. We verify that the scatter of SFR can not be significantly changed even if $L_{\rm IR}$ has an unrealistically large scatter for low-mass SFGs. Given that the combination of IR and UV luminosities is a measure of SFR, the scatter of either the IR or the UV luminosity is expected to be wider than the spread of SFR because dust extinction adds additional dispersion. 

The scatter in SSFR reflects the variation in SFH, which involves gas net accretion (inflow minus outflow) and physical processes triggering and quenching star formation, for a population of SFGs. These processes are also responsible for the broad correlations between galaxy properties (e.g., surface density of star formation and S\'{e}rsic index) and the distance off the mass-SFR relation \citep[see][for a detailed analysis]{Wuyts11b}. The observed scatter in SSFR are mostly caused by the intrinsic differences in SSFR associated with galaxy properties (e.g., color and morphology) and marginally contributed by observational errors \citep[][]{Salmi12}.
Apparently, the enhancement/quenching of star formation by violent processes such as interactions/major mergers tends to broaden the SSFR distribution and results in a larger scatter. A small scatter in SSFR, on the other hand, implies that SFGs have a similar SFH regulated by smooth/secular processes. 
The fact that the scatter in SSFR increases with stellar mass suggests that the evolutionary paradigm for SFGs is not universal along the main sequence: the dominant physical processes regulating star formation become systematically less violent in  lower-mass SFGs. This is supported by theoretical models that galaxy merger rate increases not only with redshift but also with stellar mass \citep[e.g.,][]{Hopkins10a}. 

The finding of an increasing scatter in SSFR with galaxy stellar mass is {broadly} consistent with our understanding of galaxy evolution. Local galaxies are separated by a transition stellar mass of $\log(M/M_\odot)$=\,10.5 into blue and red populations of distinct SFHs \citep{Kauffmann03}. The buildup of the red population is mostly contributed by transformation of SFGs through quenching star formation and likely morphological transition \citep{Bell07,Bundy10,Ilbert10,Bell12}. And major merger is believed to play a crucial role in driving the evolution of massive galaxies at least at $z \la 1$ but not a key mechanism for the mass assembly for low-mass SFGs \citep{Bundy09,Hopkins10a, Xu12}. Instead, environment quenching is suggested to be the major process for transformation of the low-mass SFGs into the red population at $z \la 0.5$ \citep{Peng10}. 
We notice that SSFR scatters tend to be distinct above and below the transition stellar mass $\log(M/M_\odot) \sim$ 10.5. The mass bin of $\log (M/M_\odot)$=\,10$-$10.5 exhibits the most prominent excess of outliers at the high-SSFR end from a log-normal distribution. We argue that this feature reflects a mixture of violent (e.g., interactions/major mergers) and smooth processes for regulating star formation in this mass regime. 

The outliers in the 10$^{10-10.5}\,M_\odot$ bin of the mass-SFR relation are starbursts satisfying ${\rm SSFR}>0.3$\,Gyr$^{-1}$. Taking this cut as the selection criterion for starbursts at $z\sim0.7$, we obtain that starbursts represent 7, 10, 18 and 19 percent of SFG population and contribute 19\%, 27\%, 40\% and 40\% of the total SFR in the four equally spaced mass bins from 10$^{9.5}$ to 10$^{11.5}\,M_\odot$, respectively. Mechanisms igniting starbursts are not the major process driving the evolution of SFGs. Instead, the bulk of growth through star formation is associated with the non-burst mode of star formation likely driven by smooth processes in the epoch we examine. This is consistent with the results from direct measurements of SFRs in close pairs and mergers \citep{Jogee09, Robaina09,Kaviraj13}.

\acknowledgments
We thank the anonymous referee for his/her invaluable suggestions, which have significantly improved our manuscript.
We are grateful to the COSMOS team for providing the excellent multi-wavelength data through the NASA/IPAC Infrared Science Archive, which is operated by the Jet Propulsion Laboratory, California Institute of Technology, under contract with the National Aeronautics and Space Administration.
This work is supported by National Basic Research Program of China (973 Program 2013CB834900).

\clearpage

\end{document}